\documentclass[manuscript]{emulateapj}

\def\Zsol{\hbox{Z$_{\odot}$}}

\usepackage{epsfig,natbib}
\usepackage{color}
\usepackage{soul} 
\usepackage{fancyvrb}
\usepackage{pifont}
\usepackage{longtable}
\newcommand{\oi}{O~{\sc i}}
\newcommand{\pI}{P~{\sc i}}
\newcommand{\pii}{P~{\sc ii}}
\newcommand{\piii}{P~{\sc iii}}
\newcommand{\sii}{S~{\sc ii}}
\newcommand{\hi}{H\,{\sc i}}
\newcommand{\hii}{H~{\sc ii}}
\newcommand{\Lya}{Ly$\alpha$}
\shorttitle{Tackling the Saturation of Oxygen}
\shortauthors{James \&\ Aloisi}

\begin{document}


\title{Tackling the saturation of oxygen: the use of phosphorus and sulphur as proxies within the Neutral ISM of Star-Forming Galaxies}


\author{B. James\altaffilmark{1,2 } and A. Aloisi\altaffilmark{2}}

\affil{1. Institute of Astronomy, Madingley Road, Cambridge, CB3 0HA, UK}
\affil{2. Space Telescope Science Institute, Baltimore, MD 21218}



\begin{abstract}
The abundance of oxygen in galaxies is widely used in furthering our understanding of galaxy formation and evolution. Unfortunately, direct measurements of O/H in the neutral gas are extremely difficult to obtain as the only \oi\, line available within the HST UV wavelength range (1150--3200~\AA) is often saturated. As such, proxies for oxygen are needed to indirectly derive an O/H via the assumption that solar ratios based on local Milky Way sight lines hold in different environments. In this paper we assess the validity of using two such proxies, \pii\, and \sii, within more typical star-forming environments. Using HST-COS FUV spectra of a sample of nearby star-forming galaxies and the oxygen abundances in their ionized gas, we demonstrate that both P and S are mildly depleted with respect to O and follow a trend,  $\log$(\pii/\sii)$\,=\,-1.73\,\pm\,0.18$, in excellent agreement with the solar ratio of $\log$(P/S)$_\odot=\,-1.71\,\pm\,0.04$ over the large range of metallicities (0.03--3.2~\Zsol) and \hi\ column densities ($\log[N$(\hi)/cm$^{-2}$]\,=\,18.44--21.28) spanned by the sample. From literature data we show evidence that both elements individually trace oxygen according to their respective solar ratios across a wide range of environments. Our findings demonstrate that the solar ratios of $\log$(P/O)$_\odot=-3.28\pm0.06$ and $\log$(S/O)$_\odot=-1.57\pm0.06$ can both be used to derive reliable O/H abundances in the neutral gas of local and high-redshift star-forming galaxies. The difference between O/H in the ionized- and neutral-gas phases is studied with respect to metallicity and \hi\, content. The observed trends are consistent with galactic outflows and/or star-formation inefficiency affecting the most metal-poor galaxies, with the possibility of primordial gas accretion at all metallicities.
\end{abstract}


\section{Introduction}
Oxygen is one of the most abundant elements in the Universe and is of prime importance in understanding the chemical evolution of galaxies.  Its nucleosynthesis is well known and its abundance is therefore widely used to estimate the metallicity of different Galactic and extragalactic ISM gas phases.  In particular, oxygen has been extensively investigated to derive \hii\, region metallicities as it has several strong emission lines in the optical rest frame.  However, constraining the oxygen content in the \textit{neutral} gas of a galaxy through UV absorption line studies can be challenging. Within the UV wavelength range covered by HST ($\sim$1150--3200~\AA), at low redshift the most easily observed \oi\, absorption lines are at 1302~\AA\, and 1355~\AA; the former is usually heavily saturated whilst the latter is usually too weak to be seen.

\begin{table*}
\begin{center}
\begin{scriptsize}
\caption{Logarithmic Column Densities of \hi, \oi, \pii, \& \sii\, within the Sample} \label{tab:Ndens}
%\title{Properties of Sample}
\begin{tabular}{lcccc}
\tableline\tableline
Galaxy	&	log[$N$(\hi)]			&	log[$N$(\oi)]			&	log[$N$(\pii)]			&	log[$N$(\sii)]			\\
 %& /cm$^{-2}$  & /cm$^{-2}$  & /cm$^{-2}$  & /cm$^{-2}$ \\
\tableline
I~Zw~18	&	21.28	$\pm$	0.03	& $>$	14.80	&	\ldots			&	14.72	$\pm$	0.05	\\
SBS~0335$-$052	&	21.70	$\pm$	0.05	& $>$	14.77	&	12.96	$\pm$	0.15	&	14.96	$\pm$	0.03	\\
SBS~1415+437	&	21.09	$\pm$	0.03	& $>$	14.87	&	13.41	$\pm$	0.08	&	14.94	$\pm$	0.03	\\
NGC~4214	&	21.12	$\pm$	0.03	& $>$	15.25	&	13.74	$\pm$	0.04	&	15.55	$\pm$	0.03	\\
NGC~5253-Pos.1	&	21.20	$\pm$	0.01	& $>$	15.44	&	13.73	$\pm$	0.08	&	15.43	$\pm$	0.02	\\
NGC~5253-Pos.2	&	20.65	$\pm$	0.05	& $>$	15.31	&	13.60	$\pm$	0.12	&	15.33	$\pm$	0.04	\\
NGC~4670	&	21.07	$\pm$	0.08	& $>$	15.26	&	13.80	$\pm$	0.05	& $>$	15.58			\\
NGC~4449	&	21.14	$\pm$	0.03	& $>$	15.54	&	14.09	$\pm$	0.09	&	15.60	$\pm$	0.09	\\
NGC~3690$_{v1}$	&	20.62	$\pm$	0.02	& $>$	15.48	&	\ldots			&	15.39	$\pm$	0.05	\\
NGC~3690$_{v2}$	&	19.81	$\pm$	0.08	& $>$	14.84	&	\ldots			&	\ldots			\\
M83-Pos.1	&	19.59	$\pm$	0.32	&	\ldots	&	13.84	$\pm$	0.26	&	15.39	$\pm$	0.25	\\
M83-Pos.2	& $<$	18.44			&	\ldots	& $<$	13.73			& $<$	15.39			\\\tableline
\end{tabular}
\\[1.5mm]
Notes:  All column densities refer to ICF-corrected values. 
\end{scriptsize}
\end{center}
\end{table*}

This challenging scenario often forces us to rely on a selection of elements that can, in principle, be used as proxies for oxygen.   The most suited elements for this purpose are the ones that can trace oxygen, i.e., elements that (1) deplete onto dust in a similar way to oxygen, and (2) have the same nucleosynthetic origin as oxygen.  The extent of an element's depletion onto dust grains is largely governed by the condensation temperature of that element.  Oxygen (along with carbon and nitrogen) has a relatively low condensation temperature  \citep[$T_c=182$~K,][]{Lodders:2003}which prevents it from depleting easily onto dust grains.  Conversely, refractory elements \citep[e.g., silicon with $T_c=1529$~K,][]{Lodders:2003} can be largely depleted onto dust grains. Suitable oxygen-proxy candidates should therefore have a low condensation temperature and also be $\alpha$ elements, i.e., those that arise from the same $\alpha$-capture production processes as oxygen. 

The element traditionally used as a proxy to oxygen is sulphur  \citep[e.g.,][]{Pettini:2002a,Battisti:2012,BergT:2013}.  Below 3200~\AA, the singly ionized form of sulphur (\sii) is detected through a triplet at 1250.6, 1253.8, and 1259.5~\AA. Like oxygen, sulphur has a relatively low condensation temperature \citep[$T_c=704$~K,][]{Lodders:2003} and is an $\alpha$ element (however see also \citet{Matteucci:2005} for possible non-negligible production of S by SNe Ia). While oxygen shows mild dust depletion characteristics \citep{Savage:1996}, we are still unaware as to the true refractory nature of sulphur \citep[see ][for a review]{Calura:2009ApJ}. In some environments, e.g. the cold ISM of the Milky Way, sulphur has a depletion of 0.0--0.2~dex \citep{Sembach:2000}, while models of dark interstellar clouds suggest higher depletions by 2--3 orders of magnitude \citep{Scappini:2003}.  However, since dust depletion can strongly depend on the star-formation history of a system, the depletion level of sulphur in star-forming galaxies is currently unknown and we hope to address it in this paper.  In addition to this, ionization corrections for \sii\ are found to be negligible when $\log[N$(\hi)/cm$^{-2}]\gtrsim20.5$ and gas metallicities are sub-solar \citep[][J14 hereafter]{James:2014b}. \sii\ can, however, suffer from saturation.  In a recent study of 243 sight lines in the Milky Way (MW), \citet{Jenkins:2009} (J09 hereafter) found only 9 unsaturated \sii\ lines. 

There is one other suitable oxygen-proxy candidate at our disposal within the UV wavelength range: phosphorus.  The dominant ionization stage of P in the neutral medium is \pii.  Several \pii\, lines lie within the HST UV wavelength range, the strongest being at 1152~\AA\, and 1301~\AA. However phosphorus is not an $\alpha$ element, unlike oxygen and sulphur. It is believed to be a neutron-capture element, probably formed in the carbon and neon burning shells during the late stages of the evolution of those same massive stars that form $\alpha$ elements \citep{Arnett:1996,Cescutti:2012}. The phosphorus produced in this way is then released by the explosion of these massive stars as type II SNe. According to \citet{Woosley:1995}, no significant amount of phosphorus is expected to be synthesized during the explosive phases. Phosphorus' relatively high condensation temperature \citep[$T_c=1248$~K,][]{Lodders:2003} suggests that it should be more heavily depleted.  Despite this, in studies of the ISM towards stars within the MW it has been shown that \pii\, is not depleted along sight lines containing predominantly warm low-density neutral gas \citep{Jenkins:1986} or diffuse neutral gas \citep[][L05 hereafter]{Lebouteiller:2005}. Using ten unsaturated \pii\, measurements towards local stars combined with several Galactic and extragalactic sight lines (all with $\log[N$(\hi)/cm$^{-2}]\,\geq\,20.94$), L05 found that \pii\, and \oi\, column densities relate to each other in solar P/O proportions (i.e., no differential depletion of \pii\, and \oi), suggesting that phosphorus could be a suitable tracer of oxygen in extragalactic regions. Contrary to this, J09 found that phosphorus depletes onto dust grains more rapidly than oxygen in the MW, a more metal-rich environment.  With regards to stellar abundances, there have been very few measurements for phosphorus, namely because \pI\, or \pii\, lines are absent in the ordinary wavelength range of stellar spectra.  However, in a recent study using high-resolution infrared spectroscopy of cool stars in the Galactic disk, \citet{Caffau:2011} found that the ratio between phosphorus and sulphur is roughly constant with metallicity and consistent with the solar value \citep[see also][for a more recent analysis of phosphorus abundances from the NUV spectra of a sample of 13 more metal-poor cool stars]{Jacobson:2014}.\looseness=-2

Since the majority of data concerning the relative depletions of S and P with respect to O, arises from sight lines within the local ISM, it is unclear whether these relationships only hold within similar environments, i.e., intermediate--solar metallicities\footnote{We define metallicity as 12+log(O/H), where O/H refers to the oxygen abundance of the ionized gas. In this scale the solar metallicity corresponds to 12+log(O/H)\,=\,$8.69\pm0.05$ \citep[][A09 hereafter]{Asplund:2009}. } and relatively high column densities ($\log[N$(\hi)/cm$^{-2}]\,\gtrsim\,$20).  This of course presents a serious problem as we enter an era where the improved sensitivity of instruments enables increasing access to the high-resolution rest-frame UV spectra of high-$z$ galaxies, and the need to invoke proxies for oxygen in environments that are rather different to the Milky Way becomes more frequent. Moreover, if we were to compare chemical abundances across different gas \textit{phases}, we would most benefit by constraining O/H from the neutral gas absorption line profiles because of the  common use of oxygen as metallicity indicator within the nearby universe. As of now, when comparing metallicities across cosmic time, high-$z$ observations are often forced to rely on elements that have a different nucleosynthetic origin to oxygen \citep[e.g., Fe/H,][]{Lilly:2003}.   In this paper we aim to offer a solution to the problem of assessing oxygen abundances in the neutral gas by demonstrating the reliability of sulphur and phosphorus as proxies for oxygen across a range of galaxy environments.\looseness=-2

\begin{figure*}
\begin{center}
\includegraphics[angle=0,scale=1.0]{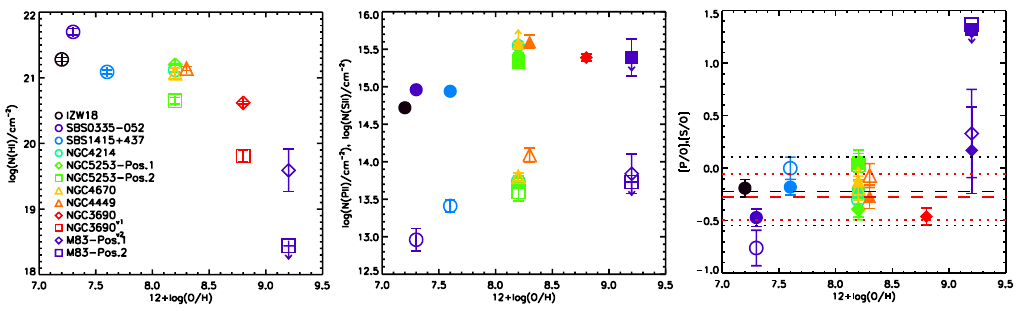}
\end{center}
\caption{\textit{Left panel:} metallicity vs. $N$(\hi); \textit{middle-panel:} metallicity vs.~$N$(\pii) and $N$(\sii) (open and filled symbols); \textit{right-panel} metallicity vs.~[P/O] and [S/O] (open and filled symbols), where [X/O]\,=\,log(X/O)\,$-$\,log(X/O)$_\odot$ and oxygen abundances refer to the ionized-gas metallicities listed in Table~\ref{tab:Oabund}. Black and red dashed lines represent weighted mean ratios of [P/O]$=-0.22\pm0.33$ and [S/O]$=-0.27\pm0.22$, respectively (excluding limits), and dotted lines represent the $\pm1\sigma$ values.} 

\label{fig:met_PSHdens}
\end{figure*}

\section{Data Analysis (an overview)}
The data utilized within this paper originate from a series of publications regarding an investigation of nearby star-forming galaxies (SFGs) using \textit{HST}-COS (PID: 11579, PI: Aloisi).  The project involved 26 orbits of COS spectroscopy with the G130M grism at $\lambda_c$=1291~\AA.  The sample of SFGs contains a range of galaxy types (i.e., blue compact dwarfs, mergers, and spirals), with oxygen metallicities of $\sim$\,0.03--3.2~\Zsol\, and distances of $\sim$\,3--54~Mpc.  Full details of the targets, observations, data reduction, and analysis techniques can be found in J14. \looseness=-2

Table~\ref{tab:Ndens} lists the column densities of \hi, \oi, \pii, and \sii\, measured within each galaxy's spectrum.  \hi\, column densities were measured from the \Lya\, profile and are in the range $\log[N$(\hi)/cm$^{-2}$]$\,\sim$\,18.4--21.7\,.  Overall, column densities for \oi\, and \pii\, were measured using \oi~$\lambda$1302 and \pii~$\lambda$1152, while for \sii\ the triplet $\lambda\lambda\lambda$1250, 1253  and 1259 was used.   A description of the specific lines used for each galaxy and pointing can be found in J14, along with their corresponding line profiles and a full description of the assessment of saturation.\looseness=-2

Ad-hoc photoionization models specific to the metallicity and \hi\ column density of each galaxy were used by J14 to estimate ionization correction factors (ICFs) to account for both classical ionization and contamination of ionized gas along the line of sight.   For \oi, \sii, and \pii, ICFs were found to be negligible for almost all galaxies, i.e., each of these ions was found to be the dominant ionization stage of the species within the neutral gas and was found not to be present within the ionized gas.  The two exceptions are M83-Pos.1 and NGC~3690$_{v2}$, where ICF corrections are $\sim$\,0.15--0.24\,dex for both \pii\, and \sii\,, and $\sim$\,0.01--0.03\,dex for \oi, mostly due to their high metallicities and the relatively low \hi\, column densities.  All column densities listed in Table~\ref{tab:Ndens} and used hereafter have been corrected for ionization using the total ICF values listed in J14 (their Table 8).  Model ICF values were unobtainable for M83-Pos.2 due to its exceptionally low \hi\ column density, and we therefore use only upper limit column densities throughout.  \looseness=-2

\section{Results}

\subsection{Column density trends with metallicity}\label{sec:trends}

The first two panels of Fig.~\ref{fig:met_PSHdens} show column densities of \hi, \sii, and \pii\, against metallicity (Table~\ref{tab:Oabund}).  These plots illustrate the wide range in metallicity and \hi\, column density covered by our sample, and show that clear relationships exist between the metallicity and column density of each species.

Firstly, the anti-correlation between metallicity and $N$(\hi) is clearly apparent: the object with the lowest \hi\, column density (M83) lies at $\sim\,3$ \Zsol\, and the objects with the highest \hi\, column density (I~Zw~18 and SBS~0335$-$052) lie well below \Zsol. 
Secondly, as the metallicity of the ionized gas increases, so do the column densities of the \sii\, and \pii\, ions up until 12+log(O/H)$\sim$8 before reaching a plateu.

The final panel of Fig.~\ref{fig:met_PSHdens} shows the [P/O] and [S/O] abundance ratios in the neutral gas against metallicities in the ionized gas. The O abundances of the neutral gas have been taken from the ionized gas metallicities given in Table~\ref{tab:Oabund} assuming that the two gas phases are homogeneus.  Since O, P, and the majority of S are all produced by the same massive stars \citep{Jacobson:2014}, and ionization corrections are minimal for \pii\ and \sii, this plot sheds light on the refractory nature of P and S.  It can be seen that for the majority of cases, both elements show mild depletion (no depletion is also consistent within the errors) relative to O (which is a mildly depleted element), with a combined average of $-0.25\pm0.39$~dex (upper limits excluded) and no significant trend with metallicity, suggesting that both P and S are mildly refractory elements in star-forming systems. The two exceptions to this are SBS~0335$-$052 and M~83, two systems at the extreme ends of metallicity and \hi\ column density covered by our sample. 

The increased [P/O] and [S/O] ratios in M~83 may signify that the elements in question do have differential depletion effects, which become more apparent at high-metallicity due to an increased dust-to-metal ratio. This effect has been seen in Damped Lyman Alpha systems (DLAs) \citep{DeCia:2013} and super-solar galaxies \citep{Brinchmann:2013}, where the dust-to-metal ratio increases as a function of metallicity. Whilst on the other hand, $\gamma$-ray burst afterglow hosts appear to contradict both these findings, with metallicity-independent and constant dust-to-metal ratios \citep{Zafar:2013}. If the former was indeed true here, and if O depletes more easily onto dust grains than P or S, then we would expect increased ratios for M~83, and perhaps even decreased ratios for SBS~0335$-$052. As such, studies of the kind presented here can, in theory, help provide insight into this controversial topic by extending the number of targets at both ends of the metallicity scale and assessing the role of differential depletion between the elements. However, it should also be noted that calculations of ionised gas abundances at high-metallicity using the direct method (as is the case here for M83) are very difficult and consequently have uncertainties which can be largely underestimated.  Given this fact, we feel that the apparent increase in depletion at high-metallicity is not significant within the uncertainties and should not be over-interpreted in this particular case.

\subsection{A solar relationship?}
\begin{figure*}
\begin{center}
\includegraphics[angle=90,scale=0.65]{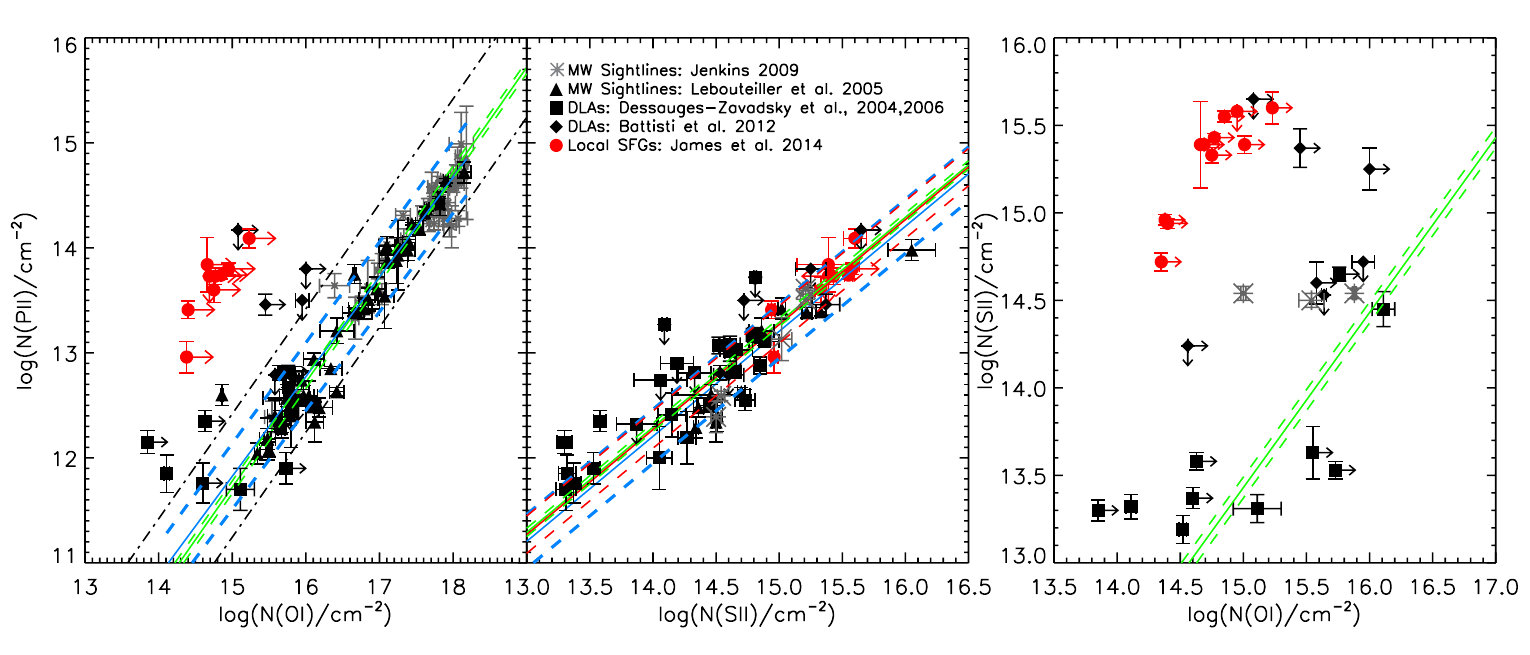}
\end{center}
\caption{Relationship between \pii, \oi, and \sii\ column densities within local SFGs (red dots; J14). Also shown (black symbols) are those measured from MW sight lines \citep{Jenkins:2009,Lebouteiller:2005} and DLAs \citep{Dessauges-Zavadsky:2004,Dessauges-Zavadsky:2006,Battisti:2012}. \textit{Left panel:} blue line represents weighted mean ratio between the compilation of \oi\ and \pii\ column densities (excluding limits) of log(\pii/\oi)\,=\,$-3.30\,\pm\,0.24$; green line represents solar ratio of log(P/O)$_\odot=-3.28\,\pm\,0.06$ (A09); black dot-dash lines represent upper and lower limits of log(P/O) derived from maximum and minimum depletions along the sight line in J09.   \textit{Middle panel:} red line corresponds to weighted mean ratio of log(\pii/\sii)$\,=\,-1.73\,\pm\,0.18$ measured from the J14 sample; blue line represents weighted mean ratio of $\log$(\pii/\sii)$\,=\,-1.80\,\pm\,0.26$ from literature values; green line represents solar ratio of log(P/S)$_\odot=-1.71\pm0.04$ (A09). \textit{Right panel:} green line represents solar ratio of log(S/O)$_\odot=-1.57\pm0.06$ (A09). In each case, dashed lines represent the $\pm1\sigma$ values.} \looseness=-2
\label{fig:P_OS}
\end{figure*}

As discussed previously, the use of certain elements as tracers for oxygen stems from the relationships observed between those elements and oxygen mostly within the Galactic neighborhood. In order to directly see one of these relationships, in the left panel of Fig.~\ref{fig:P_OS} we plot the column densities of \pii\, as a function of \oi\, using values from J09\footnote{Whilst the lines used for $N$(\pii) and $N$(\sii) are not explicitly listed by J09, strict censorship rules were employed to ensure that saturated lines were not included within the sample.} (measured from 243 sight lines within the MW) and from L05 (towards Galactic stars measured within their study or compiled from the literature). \looseness=-2

Also shown in the same panel are the \oi\, and \pii\, column densities measured for local ($0.083 < z_{abs} < 0.321$) DLAs from \citet{Battisti:2012}\footnote{For each comparison sample, column densities resulting from saturated profiles are shown as lower limits.  For the \citet{Battisti:2012} sample, upper limits refer to undetected lines and were derived from a $3\sigma$ upper limit on the equivalent width, assuming the line is on the linear part of the curve of growth.}, and high-$z$ DLAs from \cite{Dessauges-Zavadsky:2004,Dessauges-Zavadsky:2006}. Despite the fact that our literature O and P data compilation covers different environments, ranging from the local solar-metallicity neighborhood of MW sight lines, to metal-poor high-$z$ DLA systems, it can be seen that all points lie roughly along the solar P/O ratio\footnote{J09 found that in fact phosphorus depletes onto dust grains more rapidly than oxygen, as demonstrated in their paper by an oxygen depletion slope that is considerably less than that of phosphorus.  However, the left panel of Fig.~\ref{fig:P_OS} shows that P and O are still in a solar ratio in the MW sight lines probed by J09.}.  From these literature data, and excluding those values that are limits, we find the error-weighted mean ratio of $\log\,$(\pii/\oi)\,$=\,-3.30\pm0.24$ (blue solid line with the $\pm\,1\sigma$ errors indicated as blue dashed lines), which is in good agreement with the solar ratio of $\log\,$(P/O)$_{\odot}\,=\,-3.28\pm0.06$ (green solid and dashed lines). Data from our sample (J14) are also plotted in the left panel of Fig.~\ref{fig:P_OS} (red solid circles). However, due to the lower limits of the \oi\ column densities, it was not possible to calculate a mean ratio from this sample.\looseness=-2

We also determined whether \pii\, and \sii\, follow a solar ratio within local SFGs.  
The middle panel of Fig.~\ref{fig:P_OS} shows the column density of \pii\, versus that of \sii\, within our sample (red solid circles) along with the error-weighted mean column density ratio of $\log\,$(\pii/\sii)$\,=\,-1.73\pm0.18$ (red solid and dashed lines), which is in good agreement with the solar ratio of $\log$\,(P/S)$_\odot$ = $-1.71\pm0.04$.  Regarding dust depletion, the fact that P and S trace each other within these star-forming systems also suggests that both elements have similar refractory properties over a wide range of metallicities and \hi\, column densities. Indeed, we have seen from Fig.~\ref{fig:met_PSHdens} that both of these elements have a similar behaviour compared to the mildly depleted oxygen, i.e., both P and S show a mild affinity to dust.\looseness=-2

Also included in the middle panel of Fig.~\ref{fig:P_OS}  are the \sii\ and \pii\ column densities of the local and extra-galactic ISM by L05 as well as a couple of MW sight lines from J09 (due to the high-column densities involved, \sii\ was saturated in most targets of this sample), along with high-redshift \citep{Dessauges-Zavadsky:2004,Dessauges-Zavadsky:2006} and low-redshift \citep{Battisti:2012} DLAs.  When we include all (non-limit) measurements from the literature, we find a mean ratio of $\log\,$(\pii/\sii)\,$=\,-1.80\pm0.26$, which is again in agreement with the solar ratio\footnote{For reference, J09 found the slope of sulphur depletion to be in agreement with that of phosphorus within the uncertainties.}.  This result is in agreement with \citet{Caffau:2011}, where the [P/S] ratio was found to be roughly constant within 0.1~dex in 20 cool stars of the Galactic disk (see their Fig.~4), implying that P and S are produced in the same relative amounts over the $-1.0\,<\,$[Fe/H]$\,<\,+0.3$ metallicity range considered. \looseness=-2

Unfortunately, due to saturation issues affecting both S and O, the plot of $N$(\sii) versus $N$(\oi) for MW sight lines, ISM of local SFGs, and DLAs, is rather unavailing (right panel of Fig.~\ref{fig:P_OS}).  However, we can assess the relationship between S and O in stars with $T\sim\,$5000--6500\,K \citep{Nissen:2007,BergT:2013}, and ionized gas within local metal-poor SFGs \citep{Izotov:2006,Lopez-Sanchez:2010} and spiral galaxies \citep[e.g.,][]{Berg:2013}. This relationship is found to be solar in each of these studies  ($\log(S/O)_\odot=-1.57\pm0.06$).  This is of course as expected since both S and O are $\alpha$-capture elements and therefore evolve in parallel in the ISM from the time they are produced in stars until they are ejected and well mixed into the ISM. \looseness=-2

\begin{table*}
\begin{center}
\begin{scriptsize}
\caption{Oxygen abundances derived from phosphorus and sulphur column densities} \label{tab:Oabund}
\begin{tabular}{lccccccc}
\tableline\tableline
Galaxy	&	log(O$_P$/H)			&	log(O$_S$/H)			&	$\langle$log(O/H)$\rangle$		&	[O/H]$_{HI}$	&	12+log(O/H)$_{HII}$  & [O/H]$_{HII}$ & [O/H]$_{HII}-$[O/H]$_{HI}$ \\
\tableline
I~Zw~18	&		\ldots		&	$-$4.99	$\pm$	0.08	&	$-$4.99	$\pm$	0.08	&	$-$1.68	$\pm$	0.09	&	7.2 & $-$1.49  &  +0.19 $\pm$ 0.09 \\
SBS~0335$-$052	&	$-$5.46	$\pm$	0.17	&	$-$5.17	$\pm$	0.08	&	$-$5.22	$\pm$	0.07&	$-$1.92	$\pm$	0.09	&   7.3  & $-$1.39	 &  +0.53 $\pm$ 0.09 \\
SBS~1415+437	&	$-$4.40	$\pm$	0.10&	$-$4.58	$\pm$	0.07	&	$-$4.52	$\pm$	0.06&	$-$1.21	$\pm$	0.08	&   7.6  &  $-$1.09 &  +0.12 $\pm$ 0.08 \\
NGC~4214	&	$-$4.10	$\pm$	0.08	&	$-$4.00	$\pm$	0.07	&	$-$4.04	$\pm$	0.05	&	$-$0.74	$\pm$	0.07	&	8.2  &    $-$0.49	&  +0.25 $\pm$ 0.07 \\
NGC~5253-Pos.1	&	$-$4.19	$\pm$	0.10	&	$-$4.20	$\pm$	0.06	&	$-$4.20	$\pm$	0.05	&	$-$0.89	$\pm$	0.07	&  8.2  &	 $-$0.49 &  +0.40 $\pm$ 0.07 \\
NGC~5253-Pos.2	&	$-$3.77	$\pm$	0.14	&	$-$3.75	$\pm$	0.09	&	$-$3.76	$\pm$	0.08&	$-$0.45	$\pm$	0.09	&  8.2  & 	$-$0.49 &  $-$0.04 $\pm$ 0.09 \\
NGC~4670	&	$-$3.99	$\pm$	0.11	& $>$	$-$3.92			&	$-$3.99	$\pm$	0.11	&	$-$0.68	$\pm$	0.12	&  8.2 & 	$-$0.49 &  +0.19 $\pm$ 0.12 \\
NGC~4449	&	$-$3.77	$\pm$	0.11	&	$-$3.97	$\pm$	0.11	&	$-$3.87	$\pm$	0.08&	$-$0.56	$\pm$	0.09&  8.3 &  $-$0.39 &  +0.17 $\pm$ 0.09 \\
NGC~3690$_{v1}$	&		\ldots		&	$-$3.66	$\pm$	0.08	&	$-$3.66	$\pm$	0.08	&	$-$0.35	$\pm$	0.09	&  8.8 &	+0.11 &  +0.46 $\pm$ 0.09 \\
NGC~3690$_{v2}$	&	      \ldots		&		\ldots		&		\ldots		&		\ldots		&	8.8 &	+0.11 &  \ldots \\
M83-Pos.1	&	$-$2.47	$\pm$	0.42	&	$-$2.63	$\pm$	0.41	&	$-$2.55	$\pm$	0.29	&	+0.76	$\pm$	0.29& 9.2 &	+0.51 &  $-$0.25 $\pm$ 0.29 \\
M83-Pos.2	& $<$	$-$1.43			& $<$	$-$1.48			& $<$	$-$1.46			& $<$	+1.86			& 9.2 &	+0.51 &  $>$ $-$1.35 \\
\tableline
\end{tabular}
\\[1.5mm]
Notes: [O/H]\,=\,log(O/H)\,$-$\,log(O/H)$_\odot$. Average oxygen abundance [O/H]$_{HI}$ refers to weighted mean $\langle$log(O/H)$\rangle$, except for cases where only one non-limited value exists.  12+log(O/H)$_{HII}$ and [O/H]$_{HII}$ values refer to ionized-gas metallicities (see Table~1 in J14). 
\end{scriptsize}
\end{center}
\end{table*}

To summarize, \pii\, and \sii\, follow a solar ratio in our sample of nearby SFGs, similarly to what happens for a compilation of literature data representing a variety of environments.  This is somewhat expected, since sulphur and phosphorus are both products of massive stars (even though they have different nucleosynthetic origins) and since these two elements show a similar mild dust affinity in our sample of star-forming galaxies.  If we combine this result with the fact that \pii\, and \oi\, follow a solar ratio in the same literature sample, we can conclude that both \pii\ and \sii\ can indeed be used as proxies for oxygen in a variety of environments, including in the neutral gas of our nearby SFGs. \looseness=-2

\section{Discussion and Conclusions}

\begin{figure*}
\begin{center}
\includegraphics[angle=90,scale=0.8]{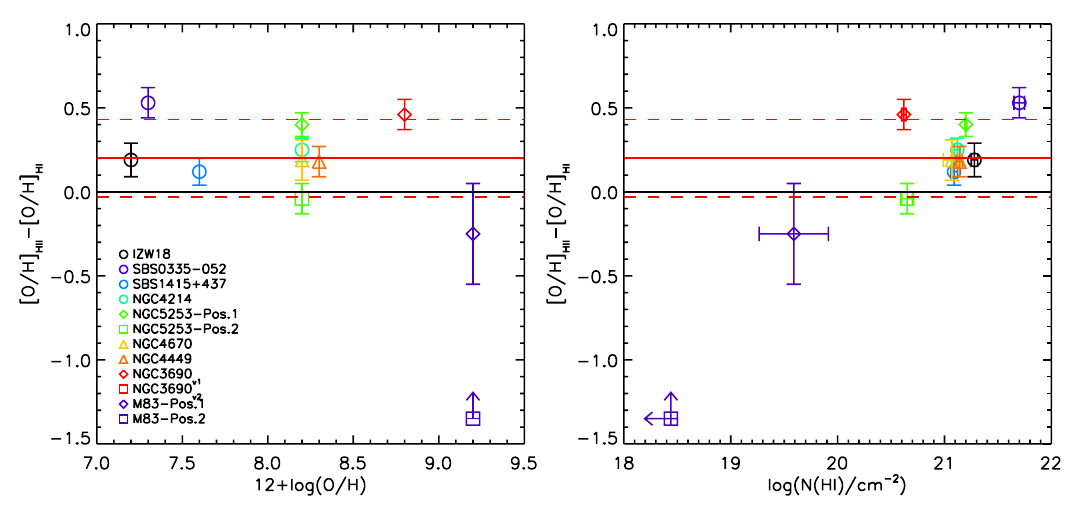}
\end{center}
\caption{Comparison between proxy-derived oxygen abundances in the neutral gas, [O/H]$_{HI}$, and ionized-gas abundances, [O/H]$_{HII}$ as a function of metallicity (\textit{left panel}) and as a function of \hi\ column density (\textit{right panel}).  Excluding limits, we find $\langle$[O/H]$_{HII}-$[O/H]$_{HI}\rangle=0.20\pm0.23$ (represented by red solid and dashed lines). Black solid line represents 1:1 ratio.}\looseness=-2
\label{fig:HII_HI}
\end{figure*}

Based on the above findings, we can now obtain O abundances for our sample of SFGs by using P and S column densities and applying the P/O and S/O solar ratios.  In Table~\ref{tab:Oabund} we list O$_P$/H and O$_S$/H which represent O/H calculated by using P and S as proxy, respectively.  We also list the mean O abundance derived from both values, $\langle$log(O/H)$\rangle$, and its value with respect to the solar abundance, [O/H]$_{HI}$.   For comparison, we also list [O/H]$_{HII}$, the ionized gas abundances from the values 12+log(O/H)$_{HII}$ measured from emission-line spectroscopy.  The last column of Table~\ref{tab:Oabund} reports the [O/H]$_{HII}-$[O/H]$_{HI}$ values as calculated from the previous quantities. Excluding limits, the mean difference between ionized- and neutral-gas abundances is $\langle$[O/H]$_{HII}-$[O/H]$_{HI}\rangle=0.20\pm0.23$. This means that by adopting the proxies here described for the neutral gas of our SFGs we can recover the O abundances of their ionized gas within a factor of 2, independently of whether the scatter is intrinsic to the sample or not.  The small scatter in the distribution of the difference between the two gas phase abundances can be clearly seen in the left-hand panel of Fig.~\ref{fig:HII_HI}, where we plot [O/H]$_{HII}-$[O/H]$_{HI}$ as a function of metallicity.  Such small offsets in the O abundances of the neutral and ionized gas of our sampled SFGs should not be taken for granted, but rather as indication that the gas appears to be relatively well mixed across a wide range in gas metallicities.  For an in-depth discussion of this issue, we refer the reader to Paper II of our COS analysis series (Aloisi et al., in prep).\looseness=-2

In the right-hand panel of Fig.~\ref{fig:HII_HI} we explore the O abundance offset further by plotting [O/H]$_{HII}-$[O/H]$_{HI}$ as a function of \hi\ column density. Here there appears to be a positive relationship between increasing $N$(\hi) and the abundance offset between the two phases.  This is in-fact a reflection of the trends shown in Fig.~\ref{fig:met_PSHdens}, where in systems with 12+log(O/H)$\gtrsim$8, the amount of P and S (and therefore O) does not change, while there is a strong decrease in \hi, thereby increasing the overall value of O/H in the neutral phase.  The opposite can also be said for low-metallicity systems, where the combination of a decrease in P, S, and therefore O and an increase in $N$(\hi), leads to lower values of O/H in the neutral phase. 

As such, Fig.~\ref{fig:HII_HI} may help shed light on the origin of metal-poor gas within low-mass, star-forming systems.  Several physical processes have been suggested and employed by chemical evolution models, to explain/achieve metal-poor gas content within such systems: (i) increasing the rate of pristine gas accretion \citep{Dalcanton:2007}; (ii) an increased efficiency of metal ejection \citep[e.g.][]{Dekel:1986,Spitoni:2010} via strong metal-rich outflows from their low gravitational potential; or (iii) having low-mass galaxies form their stars 
more slowly and therefore enrich their ISM at a slower rate than high mass galaxies \citep[i.e. `downsizing', see e.g.][]{Thomas:2005,Recchi:2009}. We discuss each of the scenarios in the following paragraphs:

Infalling \hi\, gas has been observed in several metal-poor dwarf galaxies \citep[e.g.][and references therein]{Lelli:2014,Sanchez-Almeida:2014} and is an important mechanism for triggering star-formation within these systems and causing chemical inhomogeneities. However, it is often difficult to distinguish whether the extended \hi\, structures are due to cold gas accretion from the intergalactic medium or because of a recent interaction/merger. Indeed, \citet{Lelli:2012} and \citet{Lelli:2014} interpret the \hi\, morphology of I~Zw~18 as being due to an interaction/merger between two gas-rich dwarfs rather than \hi\, gas accretion. 

Metal enhanced outflows have been confirmed both theoretically \citep[e.g.,][]{Recchi:2014} and observationally \citep[e.g.,][]{Martin:2002}. Such galactic winds in low-mass galaxies can have a large effect due to their shallow gravitational potential wells, removing freshly produced metals and leaving behind a unenriched ISM, in both the neutral and ionized phases \citep[e.g.,][]{DErcole:1999}.

Downsizing, where low-mass (and hence low-metallicity) galaxies have a lower star-formation efficiency (SFE) than high-mass systems (where SFE=SFR/M$_{gas}$, where M$_{gas}=M_{H\sc{I}}+M_{H_2}$ and SFR=star-formation rate), has been suggested by studies such as \citet{Lequeux:1979,Matteucci:1994,Calura:2009}. In this scenario, a smaller fraction of their gas has been turned into stars as a function of time, hence the lower levels of metal enrichment and larger reservoirs of \hi\, gas in low mass systems. Indeed, the effects of less efficient star-formation are commonly seen in the stellar abundances of local dwarf galaxies, where the decline in the $\alpha$/Fe ratio due to Type Ia SNe starts at lower metallicities than large spiral galaxies such as the Milky Way, e.g. \citep{Tolstoy:2009,Skuladottir:2015}.

Upon inspection of Fig.~\ref{fig:met_PSHdens} it appears that all scenarios are at play within these systems and whilst we cannot favor one fully over another, we may be able to discern if some processes are more effective than others at certain metallicities. Firstly, from the left panel we see that the column density of HI changes by $\sim$4~dex continuously as a function of metallicity over the range of metallicities sampled here.  On the other hand, the middle panel of Fig.~\ref{fig:met_PSHdens} shows that the column densities of S and P increase as a function of metallicity up until 12+log(O/H)$\sim$8, where they then plateau, with an overall change of $\lesssim$0.5\,dex and $\lesssim$1\,dex, respectively. In other words, the most metal rich galaxies have the same amount of metals, despite having a lower \hi\ gas and higher total mass (as indicated by their higher metallicity), which cannot be explained by increased star-formation efficiency or galactic outflows. This suggests that if these two processes are at play, they mostly affect the most metal-poor (i.e. 12+log(O/H)$\lesssim$8), low-mass systems. This of course, could be in addition to increased accretion at low metallicity, which by itself could easily explain the decrease in \hi\ gas and increase in metallicity across the entire metallicity range (Fig.~\ref{fig:met_PSHdens}, left-panel). In order to properly disentangle these affects, we would require additional information concerning the star-formation histories and full chemical abundances of each system, which is outside the scope of this study.

As we have now established that both S and P are reliable tracers of O in the neutral ISM, a question remains as to whether circumstances exist where one proxy is more suitable than the other.  This can be evaluated via the properties of each species.  Firstly, their affinity to become saturated can be assessed using the curve of growth (COG), which plots $\log$\,($N\,\times\,f\,\times\,\lambda$) against $\log\,(W/\lambda$) (where $N$\,=\,column density, $f$\,=\,oscillator strength, $\lambda$\,=\,wavelength, and $W$\,=\,equivalent width) - i.e., the location of an absorption line along the COG can reveal whether or not the line is saturated (see, e.g., J14).  Both \pii~$\lambda$1152 and the weakest of the \sii\ triplet,  \sii~$\lambda$1250, fall on the same part of the COG.   Using the column densities of Table~\ref{tab:Ndens}, for \pii~$\lambda$1152 and \sii~$\lambda$1250 \citep[with $f=2.361\times10^{-1}$ and $f=5.453\times10^{-3}$, respectively; see][]{Morton:1991}, we obtain averages of $\log(Nf\lambda)=8.171$ and 8.174, respectively.  As they lie so close on the COG, there would be few circumstances when one would become saturated without the other. It should be noted that while \pii~$\lambda$1301 lies slightly lower on the COG ($f=1.271\times10^{-2}$, giving $\log(Nf\lambda)=6.957$ on average), the line is heavily contaminated with \oi~$\lambda$1302 and therefore unusable.  The third \pii\ line within the HST-UV wavelength range, \pii~$\lambda$1532, is $\sim$\,30 times fainter than  \pii~$\lambda$1301 and therefore not easily detected.  If only one or both of the other \sii\ triplet lines are available instead of \sii~$\lambda$1250, then there will be cases where \sii\ is saturated and \pii\ is not; e.g., this is the case for NGC~4670 in all the sight lines studies within J14.\looseness=-2

With regards to their wavelength, having only $\sim$100~\AA\ separation between \pii~$\lambda$1152 and \sii~$\lambda$1250 means that they are both readily accessible within the FUV spectrum and cases where only \sii\ or \pii\ is available would be most likely rare (unless one of the two is blended, e.g., with a MW absorption).  On the other hand, this does mean that the observer will benefit from two oxygen-proxy elements within a single spectrum, thereby providing a more accurate oxygen abundance determination.\looseness=-2

A small difference does exist between the ionization potentials of \pii\ (19.76946~eV) and \sii\ (23.33788~eV) \citep{Morton:2003}.  As such, environments may exist where the strength of the ionizing source may affect one species more than another, e.g., along sight lines that have low \hi\ column density.  In such cases, \pii\ would be more easily ionized than \sii\ and a larger portion of P would exist as \piii\, in the ionized gas.  This could of course be corrected for by applying adequate ICFs (as seen in Table~8 of J14).  \looseness=-2

Overall it can be seen that \textit{both} \pii\ and \sii\ can provide a mean for deriving accurate oxygen abundances in the neutral ISM of SFGs.  Using our sample of local SFGs (J14), we have demonstrated that both P and S show minimal depletion with respect to O, suggesting that both elements are mildly refractory elements in star-forming systems. We have also shown that P/S follows the solar ratio of $\log$(P/S)$_\odot=-1.71\pm0.04$ across the wide range of \hi\ column densities ($\log[N$(\hi)/$cm^{-2}$]\,=\,18.44--21.28) and metallicities (0.03--3.2~\Zsol) of our sample, with an average of $\log$(\pii/\sii)\,$=-1.73\pm0.18$.  We have additionally demonstrated from the literature that ratios of $\log$(\pii/\sii)\,$=-1.80\pm0.26$ and $\log$(\pii/\oi)$=-3.30\pm0.24$ for sight lines throughout a wide range of environments, such as DLAs (similar to our sample of SFGs) and the MW, are also in very good agreement with solar values ($\log$(P/O)$_\odot=-3.28\pm0.06$). By combining these three pieces of information with the fact that S/O also follows the solar ratio of $\log$(S/O)$_\odot=-1.57\pm0.06$ in stellar abundances and the ionized gas within SFGs, we suggest that both P and S can reliably trace O throughout the neutral gas of typical SFGs.\looseness=-2

In addition to this, our results show that while the column density of metals (i.e., P, S, and therefore O) remains constant in the neutral phase for galaxies with 12+log(O/H)$\gtrsim$8.0, the HI column density increases with decreasing metallicity across the whole sample, resulting in an overall decreased O/H abundance ratio in the neutral gas of systems with higher $N$(HI). This is consistent with galactic outflows and/or low star-formation efficiency preferentially affecting low-mass galaxies, whilst accretion of primordial gas may be affecting SFGs across the wider metallicity range.

By covering a large range in ISM properties, and in particular extending down to significantly low metallicity, we can ensure that the application of a P/O and S/O solar ratio to infer O from P and S can be both numerous and widespread. In particular, the derivation of neutral ISM oxygen abundances in high-$z$ systems can now be derived with a higher level of certainty, either from using P or S, or by using the solar relationships to check whether the measured O/H is based on a unsaturated line or not.  We hope the relations derived within this work will aid future studies in comparing oxygen abundances across different gas phases and enable oxygen abundance measurements throughout cosmic time.\looseness=-2

\bibliographystyle{mn2e}
\bibliography{references.bib}

\begin{thebibliography}{47}
\expandafter\ifx\csname natexlab\endcsname\relax\def\natexlab#1{#1}\fi

\bibitem[{{Arnett}(1996)}]{Arnett:1996}
{Arnett} D., 1996, {Supernovae and Nucleosynthesis: An Investigation of the
  History of Matter from the Big Bang to the Present}

\bibitem[{{Asplund} {et~al.}(2009){Asplund}, {Grevesse}, {Sauval}, \&
  {Scott}}]{Asplund:2009}
{Asplund} M., {Grevesse} N., {Sauval} A.~J., {Scott} P., 2009, \araa, 47, 481

\bibitem[{Battisti {et~al.}(2012)Battisti, Meiring, Tripp, Prochaska, Werk,
  Jenkins, Lehner, Tumlinson, \& Thom}]{Battisti:2012}
Battisti A.~J., Meiring J.~D., Tripp T.~M., Prochaska J.~X., Werk J.~K.,
  Jenkins E.~B., Lehner N., Tumlinson J., Thom C., 2012, \apj, 744, 93

\bibitem[{{Berg} {et~al.}(2013{\natexlab{a}}){Berg}, {Skillman}, {Garnett},
  {Croxall}, {Marble}, {Smith}, {Gordon}, \& {Kennicutt}}]{Berg:2013}
{Berg} D.~A., {Skillman} E.~D., {Garnett} D.~R., {Croxall} K.~V., {Marble}
  A.~R., {Smith} J.~D., {Gordon} K., {Kennicutt} Jr. R.~C., 2013{\natexlab{a}},
  \apj, 775, 128

\bibitem[{{Berg} {et~al.}(2013{\natexlab{b}}){Berg}, {Ellison}, {Venn}, \&
  {Prochaska}}]{BergT:2013}
{Berg} T.~A.~M., {Ellison} S.~L., {Venn} K.~A., {Prochaska} J.~X.,
  2013{\natexlab{b}}, \mnras, 434, 2892

\bibitem[{{Brinchmann} {et~al.}(2013){Brinchmann}, {Charlot}, {Kauffmann},
  {Heckman}, {White}, \& {Tremonti}}]{Brinchmann:2013}
{Brinchmann} J., {Charlot} S., {Kauffmann} G., {Heckman} T., {White} S.~D.~M.,
  {Tremonti} C., 2013, \mnras, 432, 2112

\bibitem[{{Caffau} {et~al.}(2011){Caffau}, {Bonifacio}, {Faraggiana}, \&
  {Steffen}}]{Caffau:2011}
{Caffau} E., {Bonifacio} P., {Faraggiana} R., {Steffen} M., 2011, \aap, 532,
  A98

\bibitem[{{Calura} {et~al.}(2009{\natexlab{a}}){Calura}, {Dessauges-Zavadski},
  {Prochaska}, \& {Matteucci}}]{Calura:2009ApJ}
{Calura} F., {Dessauges-Zavadski} M., {Prochaska} J.~X., {Matteucci} F.,
  2009{\natexlab{a}}, \apj, 693, 1236

\bibitem[{{Calura} {et~al.}(2009{\natexlab{b}}){Calura}, {Pipino}, {Chiappini},
  {Matteucci}, \& {Maiolino}}]{Calura:2009}
{Calura} F., {Pipino} A., {Chiappini} C., {Matteucci} F., {Maiolino} R.,
  2009{\natexlab{b}}, \aap, 504, 373

\bibitem[{{Cescutti} {et~al.}(2012){Cescutti}, {Matteucci}, {Caffau}, \&
  {Fran{\c c}ois}}]{Cescutti:2012}
{Cescutti} G., {Matteucci} F., {Caffau} E., {Fran{\c c}ois} P., 2012, \aap,
  540, A33

\bibitem[{{Dalcanton}(2007)}]{Dalcanton:2007}
{Dalcanton} J.~J., 2007, \apj, 658, 941

\bibitem[{{De Cia} {et~al.}(2013){De Cia}, {Ledoux}, {Savaglio}, {Schady}, \&
  {Vreeswijk}}]{DeCia:2013}
{De Cia} A., {Ledoux} C., {Savaglio} S., {Schady} P., {Vreeswijk} P.~M., 2013,
  \aap, 560, A88

\bibitem[{{Dekel} \& {Silk}(1986)}]{Dekel:1986}
{Dekel} A., {Silk} J., 1986, \apj, 303, 39

\bibitem[{{D'Ercole} \& {Brighenti}(1999)}]{DErcole:1999}
{D'Ercole} A., {Brighenti} F., 1999, \mnras, 309, 941

\bibitem[{{Dessauges-Zavadsky} {et~al.}(2004){Dessauges-Zavadsky}, {Calura},
  {Prochaska}, {D'Odorico}, \& {Matteucci}}]{Dessauges-Zavadsky:2004}
{Dessauges-Zavadsky} M., {Calura} F., {Prochaska} J.~X., {D'Odorico} S.,
  {Matteucci} F., 2004, \aap, 416, 79

\bibitem[{{Dessauges-Zavadsky} {et~al.}(2006){Dessauges-Zavadsky}, {Prochaska},
  {D'Odorico}, {Calura}, \& {Matteucci}}]{Dessauges-Zavadsky:2006}
{Dessauges-Zavadsky} M., {Prochaska} J.~X., {D'Odorico} S., {Calura} F.,
  {Matteucci} F., 2006, \aap, 445, 93

\bibitem[{{Izotov} {et~al.}(2006){Izotov}, {Stasi{\'n}ska}, {Meynet}, {Guseva},
  \& {Thuan}}]{Izotov:2006}
{Izotov} Y.~I., {Stasi{\'n}ska} G., {Meynet} G., {Guseva} N.~G., {Thuan} T.~X.,
  2006, \aap, 448, 955

\bibitem[{{Jacobson} {et~al.}(2014){Jacobson}, {Thanathibodee}, {Frebel},
  {Roederer}, {Cescutti}, \& {Matteucci}}]{Jacobson:2014}
{Jacobson} H.~R., {Thanathibodee} T., {Frebel} A., {Roederer} I.~U., {Cescutti}
  G., {Matteucci} F., 2014, \apjl, 796, L24

\bibitem[{{James} {et~al.}(2014){James}, {Aloisi}, {Heckman}, {Sohn}, \&
  {Wolfe}}]{James:2014b}
{James} B.~L., {Aloisi} A., {Heckman} T., {Sohn} S.~T., {Wolfe} M.~A., 2014,
  \apj, 795, 109

\bibitem[{{Jenkins}(1986)}]{Jenkins:1986}
{Jenkins} E.~B., 1986, \apj, 304, 739

\bibitem[{{Jenkins}(2009)}]{Jenkins:2009}
---, 2009, \apj, 700, 1299

\bibitem[{{Lebouteiller} {et~al.}(2005){Lebouteiller}, {Kuassivi}, \&
  {Ferlet}}]{Lebouteiller:2005}
{Lebouteiller} V., {Kuassivi}, {Ferlet} R., 2005, \aap, 443, 509

\bibitem[{{Lelli} {et~al.}(2014){Lelli}, {Verheijen}, \&
  {Fraternali}}]{Lelli:2014}
{Lelli} F., {Verheijen} M., {Fraternali} F., 2014, \mnras, 445, 1694

\bibitem[{{Lelli} {et~al.}(2012){Lelli}, {Verheijen}, {Fraternali}, \&
  {Sancisi}}]{Lelli:2012}
{Lelli} F., {Verheijen} M., {Fraternali} F., {Sancisi} R., 2012, \aap, 537, A72

\bibitem[{{Lequeux} {et~al.}(1979){Lequeux}, {Peimbert}, {Rayo}, {Serrano}, \&
  {Torres-Peimbert}}]{Lequeux:1979}
{Lequeux} J., {Peimbert} M., {Rayo} J.~F., {Serrano} A., {Torres-Peimbert} S.,
  1979, \aap, 80, 155

\bibitem[{{Lilly} {et~al.}(2003){Lilly}, {Carollo}, \& {Stockton}}]{Lilly:2003}
{Lilly} S.~J., {Carollo} C.~M., {Stockton} A.~N., 2003, \apj, 597, 730

\bibitem[{{Lodders}(2003)}]{Lodders:2003}
{Lodders} K., 2003, \apj, 591, 1220

\bibitem[{{L{\'o}pez-S{\'a}nchez} \& {Esteban}(2010)}]{Lopez-Sanchez:2010}
{L{\'o}pez-S{\'a}nchez} {\'A}.~R., {Esteban} C., 2010, \aap, 517, A85

\bibitem[{{Martin} {et~al.}(2002){Martin}, {Kobulnicky}, \&
  {Heckman}}]{Martin:2002}
{Martin} C.~L., {Kobulnicky} H.~A., {Heckman} T.~M., 2002, \apj, 574, 663

\bibitem[{{Matteucci}(1994)}]{Matteucci:1994}
{Matteucci} F., 1994, \aap, 288, 57

\bibitem[{{Matteucci} \& {Chiappini}(2005)}]{Matteucci:2005}
{Matteucci} F., {Chiappini} C., 2005, PASA, 22, 49

\bibitem[{{Morton}(1991)}]{Morton:1991}
{Morton} D.~C., 1991, \apjs, 77, 119

\bibitem[{{Morton}(2003)}]{Morton:2003}
---, 2003, \apjs, 149, 205

\bibitem[{{Nissen} {et~al.}(2007){Nissen}, {Akerman}, {Asplund}, {Fabbian},
  {Kerber}, {Kaufl}, \& {Pettini}}]{Nissen:2007}
{Nissen} P.~E., {Akerman} C., {Asplund} M., {Fabbian} D., {Kerber} F., {Kaufl}
  H.~U., {Pettini} M., 2007, \aap, 469, 319

\bibitem[{{Pettini} {et~al.}(2002){Pettini}, {Ellison}, {Bergeron}, \&
  {Petitjean}}]{Pettini:2002a}
{Pettini} M., {Ellison} S.~L., {Bergeron} J., {Petitjean} P., 2002, \aap, 391,
  21

\bibitem[{{Recchi}(2014)}]{Recchi:2014}
{Recchi} S., 2014, Advances in Astronomy, 2014, 750754

\bibitem[{{Recchi} {et~al.}(2009){Recchi}, {Calura}, \& {Kroupa}}]{Recchi:2009}
{Recchi} S., {Calura} F., {Kroupa} P., 2009, \aap, 499, 711

\bibitem[{{S{\'a}nchez Almeida} {et~al.}(2014){S{\'a}nchez Almeida},
  {Morales-Luis}, {Mu{\~n}oz-Tu{\~n}{\'o}n}, {Elmegreen}, {Elmegreen}, \&
  {M{\'e}ndez-Abreu}}]{Sanchez-Almeida:2014}
{S{\'a}nchez Almeida} J., {Morales-Luis} A.~B., {Mu{\~n}oz-Tu{\~n}{\'o}n} C.,
  {Elmegreen} D.~M., {Elmegreen} B.~G., {M{\'e}ndez-Abreu} J., 2014, \apj, 783,
  45

\bibitem[{{Savage} \& {Sembach}(1996)}]{Savage:1996}
{Savage} B.~D., {Sembach} K.~R., 1996, \apj, 470, 893

\bibitem[{{Scappini} {et~al.}(2003){Scappini}, {Cecchi-Pestellini}, {Smith},
  {Klemperer}, \& {Dalgarno}}]{Scappini:2003}
{Scappini} F., {Cecchi-Pestellini} C., {Smith} H., {Klemperer} W., {Dalgarno}
  A., 2003, \mnras, 341, 657

\bibitem[{{Sembach} {et~al.}(2000){Sembach}, {Howk}, {Ryans}, \&
  {Keenan}}]{Sembach:2000}
{Sembach} K.~R., {Howk} J.~C., {Ryans} R.~S.~I., {Keenan} F.~P., 2000, \apj,
  528, 310

\bibitem[{{Sk{\'u}lad{\'o}ttir} {et~al.}(2015){Sk{\'u}lad{\'o}ttir},
  {Andrievsky}, {Tolstoy}, {Hill}, {Salvadori}, {Korotin}, \&
  {Pettini}}]{Skuladottir:2015}
{Sk{\'u}lad{\'o}ttir} {\'A}., {Andrievsky} S.~M., {Tolstoy} E., {Hill} V.,
  {Salvadori} S., {Korotin} S.~A., {Pettini} M., 2015, \aap, 580, A129

\bibitem[{{Spitoni} {et~al.}(2010){Spitoni}, {Calura}, {Matteucci}, \&
  {Recchi}}]{Spitoni:2010}
{Spitoni} E., {Calura} F., {Matteucci} F., {Recchi} S., 2010, \aap, 514, A73

\bibitem[{{Thomas} {et~al.}(2005){Thomas}, {Maraston}, {Bender}, \& {Mendes de
  Oliveira}}]{Thomas:2005}
{Thomas} D., {Maraston} C., {Bender} R., {Mendes de Oliveira} C., 2005, \apj,
  621, 673

\bibitem[{{Tolstoy} {et~al.}(2009){Tolstoy}, {Hill}, \& {Tosi}}]{Tolstoy:2009}
{Tolstoy} E., {Hill} V., {Tosi} M., 2009, \araa, 47, 371

\bibitem[{{Woosley} \& {Weaver}(1995)}]{Woosley:1995}
{Woosley} S.~E., {Weaver} T.~A., 1995, \apjs, 101, 181

\bibitem[{{Zafar} \& {Watson}(2013)}]{Zafar:2013}
{Zafar} T., {Watson} D., 2013, \aap, 560, A26

\end{thebibliography}

\acknowledgments
We gratefully acknowledge Max Pettini for useful discussions and insightful comments on this manuscript. We also thank an anonymous referee for thoughtful comments that greatly improved the paper. STScI is operated by the Association of Universities for Research in Astronomy, Inc., under NASA contract NAS5-26555. Support for Program number 11579 was provided by NASA through a grant from the Space Telescope Science Institute, which is operated by the Association of Universities for Research in Astronomy, Incorporated, under NASA contract NAS5-26555.

\end{document}